\documentclass[publish,IJAT,paper]{ijatarticle}

\usepackage{graphicx}
\usepackage{epsfig}
\usepackage{amsmath}
\usepackage{amssymb}
\usepackage{algorithm,algorithmic}
\usepackage{here}
\usepackage{multicol}
\usepackage{multirow}
\usepackage{comment}
\usepackage{lscape}

\usepackage{amsthm}
\usepackage{cite}
\usepackage{subcaption}

\newcommand{\xv}{\bm{x}}
\newcommand{\xiv}{\bm{\xi}}

\theoremstyle{plain}
\newtheorem{rem}{Remark}
\newtheorem{lem}{Lemma}
\newtheorem{pf}{Proof}
\newtheorem{dfn}{Definition}

\SetVolumeNumber{Vol.0 No.0, 200x} 
\SetArticleID{Au*-*-****; \jtoday}
\SetIndexTab{}%

\begin{document}

\pagestyle{ijatstyle}
\title{Hierarchical-type Model Predictive Control and Experimental Evaluation for a Water-Hydraulic Artificial Muscle with Direct Data-Driven Adaptive Model Matching}
\author{Satoshi Tsuruhara$^{*,**}$, and Kazuhisa Ito$^{***}$}
\address{$^*$Graduate School of Engineering and Science, 
Shibaura Institute of Technology,\\
$307$ Fukasaku, Minuma, Saitama 3378570, Japan.\\
         E-mail: nb23110@shibaura-it.ac.jp\\
         $^{**}$Research fellow,\\
         Japan Society for the Promotion of Science,\\
         Tokyo 1020083, Japan.\\
         $^{***}$Department of Machinery and Control Systems,\\
         Shibaura Institute of Technology,\\
         307 Fukasaku, Minuma, Saitama 3378570, Japan.\\
         E-mail: kazu-ito@shibaura-it.ac.jp\\}
\markboth{Tsuruhara, S., Ito, K.}{Hierarchical-type Model Predictive Control and Experimental Evaluation for a Water-Hydraulic Artificial Muscle with Direct Data-Driven Adaptive Model Matching}
\dates{00/00/00}{00/00/00}
\maketitle

\begin{abstract}
\noindent High-precision displacement control for water-hydraulic artificial muscles is a challenging issue due to its strong hysteresis characteristics that is hard to be modelled precisely, and many control methods have been proposed.
Recently, data-driven control methods have attracted much attention because they do not explicitly use mathematical models, making design much easier.
In our previous work, we proposed fictitious reference iterative tuning (FRIT)-based model predictive control (FMPC), which combines data-driven and model-based methods for the muscle and showed its effectiveness because it can consider input constraints as well.
However, the problem in which control performance strongly depends on prior input-output data remains still unsolved.
Adaptive FRIT based on directional forgetting has also been proposed; however, it is difficult to achieve the desired transient performance because it cannot consider input constraints and there are no design parameters that directly determine the control performance, such as MPC.
In this study, we propose a novel data-driven adaptive model matching-based controller that combines these methods.
Experimental results show that the proposed method could significantly improve the control performance and achieve high robustness against inappropriate initial experimental data , while considering the input constraints in the design phase.
\end{abstract}

\begin{keywords}
adaptive model matching, data-driven control, model predictive control, water-hydraulic artificial muscle, displacement control 
\end{keywords}

\section{Introduction}
A water-hydraulic artificial muscle has the advantages of both McKibben-type artificial muscles and tap-water-driven systems. Artificial muscles are lightweight, low-cost, and have a high power density.
Moreover, tap-water-driven systems do not require compressors or special facilities, are low-noise, and are oil-free, resulting in a low environmental burden \cite{MBAMM}.
These advantages allow us to expand their applications to various fields, including exploration and rescue, medical and rehabilitation, industrial applications, and biomimetic robots \cite{PAM1}.
However, the muscles exhibit undesirable asymmetric hysteresis characteristics that strongly depend on the load and frequency of the target trajectory, making it difficult to design a control system \cite{MBAMM}.
Therefore, it is challenging to propose a practical control method for achieving high-precision displacement control of the muscles.

In recent years, data-driven control (DDC) have been developed because they can achieve the desired control performance without explicit mathematical model \cite{DDC1,DDC2,DDC3}.
There are several types of DDCs, such as fictitious reference iterative tuning (FRIT) \cite{FRIT} and virtual reference feedback tuning \cite{VRFT}, which are parameter-tuning methods; model-free adaptive control \cite{MFAC} or model-free control \cite{MFC} based on data models, and neural network \cite{NN} and reinforcement learning \cite{RL}, which are machine-learning-based methods.
Many of these DDCs have been applied to the muscles, and their effectiveness has been verified \cite{DDC_PAM1,DDC_PAM2,DDC_PAM3,DDC_PAM4,DDC_PAM5}. 

However, compared with model-based control methods, the number of tuning parameters to be tuned or optimized increases, and it is difficult to satisfy input constraints, such as the applied voltage at the design stage for practical systems.
To address these problems, several hierarchical control methods that combine data-driven and model-based control methods have been proposed \cite{HC1,HC2,HC3,HC4}.
These methods have a similar structure in that the inner loop is designed using data-driven methods and model-based methods are employed in the outer loop.
High-precision control can be achieved using a control system that utilizes both data-driven and model-based control.
However, their control performance is strongly dependent on prior data and its working conditions, because the control parameters of the inner loop are already fixed, or their controller cannot consider input constraints at the design stage.
The authors have also proposed a new method that consider a similar mechanism, using Extended-FRIT (E-FRIT) \cite{E-FRIT}, data-driven control, for the inner loop and MPC that considers input constraints for the outer loop \cite{FMPC1,FMPC2}.
This E-FRIT-based MPC (FMPC) was significant that guaranteed the input constraints and was independent of the design of the reference model.

As with other hierarchical controllers, the control performance of the FRIT-based MPC (FMPC) is strongly dependent on the experimental data and the weight of the input difference terms in E-FRIT.
Hence, this is challenging issue for the following reasons;
1) the characteristics of FRIT/E-FRIT, which tunes the controller based on a single prior experimental data set, 2) the optimal time constant in the reference model changes due to the weight of the input difference, which is the design parameter \cite{FMPC1}, and 3) optimization results depend on the initial controller parameters such as PID gains of the prior experiment due to nonlinear optimization problem for FRIT/E-FRIT.
Therefore, 
If inappropriate values are selected, control performance of FMPC is significantly degraded or destabilized.
Hence, the advantage of FRIT/E-FRIT, in which the controller is optimized in one shot, is diminished.

By contrast, adaptive FRIT (A-FRIT) was proposed by \cite{A-FRIT1} and extended to A-FRIT with directional forgetting (DF) to further robustness improvement by \cite{A-FRIT2}.
These results show that DF-based A-FRIT can achieve better control performance without relying on optimality of the design parameter of FMPC.
However, the transient performance is significantly degraded because this method cannot consider input constraints.

In this study, to overcome the shortcomings of both FMPC and A-FRIT methods, we proposed a novel hierarchical control method that combines DF-based A-FRIT, which serve as a data-driven adaptive model matching, with FMPC \cite{A-FMPC}.
The proposed method combines the pseudo-linearization (PL) technique using A-FRIT in the inner loop and MPC adopting this PL model as a predictor in the outer loop.
Adaptive model matching based on A-FRIT, which is data-driven method, is applied in contrast to conventional methods, which use a fixed controller for model matching to the PL model.
This is expected to enable the design of control systems relaxing the strong dependency on prior experimental data.
Moreover, unlike the traditional adaptive model predictive control \cite{AMPC}, the model used in the predictor remains fixed, allowing independent design of the inner and outer loops, which has the advantage of facilitating the design of the MPC weight parameters.

We applied the proposed method to a water-hydraulic artificial muscle and experimentally validated its effectiveness.
In particular, the influence and robustness against 1) weight of input difference for E-FRIT, 2) initial control parameters, and 3) pre-experiments that differ from experimental conditions, are discussed in many experimental data.
These results are expected to significantly contribute to the design guidelines for FMPC and A-FMPC.

The main contributions of this study are as follows.
1) A novel model predictive control system based on a data-driven adaptive model matching method using DF-based A-FRIT is proposed.
The method has the advantages of both A-FRIT and FMPC while improving the disadvantages of each.
2) Experiments were performed to clarify the relationship between design parameters of the input change for parameter tuning in offline and control performance.
3) The proposed method achieves high robustness against design parameters, initial control parameters, and prior experimental data.
The proposed method is much easier to design than conventional methods.

The remainder of this paper is organized as follows.
First, Section 2 briefly describes FRIT and its extensions, E-FRIT and A-FRIT.
Next, Section 3 describes the proposed method, which applies data-driven adaptive model matching based on A-FRIT to define a PL model.
We design an adaptive model predictive control system using the PL model as a predictor.
Section 4 presents the results of applying the proposed method to a water hydraulic artificial muscle, compares it with conventional methods, and evaluates its robustness.
Finally, Section 5 presents conclusions and future works.

\section{Overview of A-FRIT algorithm}
In this section, the FRIT algorithm is briefly presented, and a DF-based A-FRIT algorithm \cite{A-FRIT2} is explained, which makes data-driven adaptive model matching as the proposed methods.
\subsection{FRIT}
As a pioneering work, FRIT proposed by \cite{FRIT}, is a type of data-driven tuning method that can tune the controller based on single prior input-output (I/O) data and has the advantage of requiring less assumptions and no pre-filter design unlike VRFT.
Let $z$ be the time-shift operator and $k$ be the time step.
FRIT is a model-matching method that aims to minimize the following evaluation function, which is the difference between a closed-loop system and the reference model $G_m(z)$.
\begin{equation}\label{ma:MR}
J(z,\theta)=\left\|\frac{P(z)C(z,\theta)}{1+P(z)C(z,\theta)}r(k)-G_m(z)r(k)\right\|_2^2,
\end{equation}
where $\|\cdot\|_2$ represents the Euclidean norm, $P(z)$ denotes the plant, which is unknown, and $r(k)\in\mathbb{R}$ denotes the target trajectory.
In this study, the controller structure $C(z,\theta)\triangleq\theta^T\beta(z)$ is set as a discrete-time PID controller and can be expressed as follows:
\begin{equation}
\begin{split}
\theta &= [K_p,\ K_i,\ K_d]^T,\\
\beta(z)&=[1,\ {T_s}/{(1-z^{-1})},\ {(1-z^{-1})}/{T_s}]^T,
\end{split}
\end{equation}
where $T_s$ denotes the sampling time, and $K_p, K_i, K_d$ denote the PID gains.
Note that the controller can be selected arbitrarily if the control structure is linear with respect to the parameter $\theta$.
Actually, (\ref{ma:MR}) cannot be solved because the evaluation function has the unknown plant $P(z)$.
Using a prior single input-output (I/O) data $y_0(k), u_0(k)\in\mathbb{R}^N$, we transform the evaluation function (\ref{ma:MR}) into an optimization problem based on the following a fictitious reference signal $\tilde{r}(k)\in\mathbb{R}^N$,
\begin{equation}\label{ma:r_tilde}
\tilde{r}(\theta,k)=C^{-1}(z,\theta)u_0(k)+y_0(k).
\end{equation}
Then,
\begin{equation}
\theta^{*} = \mathrm{arg}\ \min_\theta\ J_{\mathrm{FRIT}}(\theta),
\end{equation}
\begin{equation}\label{ma:eva_FRIT}
J_{\mathrm{FRIT}}(\theta)=\sum_{k=1}^{N}\left[y_0(k)-G_m(z)\tilde{r}(\theta,k)\right]^2,
\end{equation}
where $\theta^{*}$ denote the optimal parameters of $\theta$.
FRIT is assumed to require no mathematical model; thus, (\ref{ma:eva_FRIT}) does not include unknown values.
Therefore, the controller can be designed using only I/O data by solving the nonlinear optimization problem.

\subsection{E-FRIT}
FRIT is a powerful method, but it may generate excessive control inputs depending on the I/O data from prior experiments and the select of reference model $G_m(z)$.
In practical application, it is extended to the minimization problem for evaluation function with the addition of a variable term in the input as follows:
\begin{equation}
\theta^{*} = \mathrm{arg}\ \min_\theta\ J_{\mathrm{E-FRIT}}(\theta),
\end{equation}
\begin{equation}\label{ma:eva_EFRIT}
J_{\mathrm{E-FRIT}}(\theta)=\sum_{k=1}^{N}\left[y_0(k)-G_m(z)\tilde{r}(\theta,k)\right]^2+\lambda\sum_{k=1}^{N}\Delta\tilde{u}^2(\theta,k),
\end{equation}
where $\Delta\tilde{u}(\theta,k)\triangleq\tilde{u}(\theta,k)-\tilde{u}(\theta,k-1)$ and $\lambda>0\in\mathbb{R}$ denote the virtual input difference and its weights, respectively.
Note that when $\lambda=0$ in evaluation function (\ref{ma:eva_EFRIT}), it reduces to the normal FRIT and the extension to A-FRIT uses it.
\begin{algorithm}[t]
    \caption{FRIT/EFRIT algorithm}\label{alg1}
    \begin{algorithmic}
        \REQUIRE {$\beta(z)$, $G_m(z)$, $\theta_0$, $\lambda>0$}
        \ENSURE {$\theta$}
        \STATE{\textbf{Step 1}: Obtain:}
        \STATE{$\qquad u_0(k),y_0(k), k=1,\cdots,N$}
        \STATE{\textbf{Step 2}: Calculate:}
        \STATE{$\qquad$ fictitious reference signal:}
        \STATE{$\qquad\quad\tilde{r}(k,\theta,\theta_0)\leftarrow C^{-1}(z,\theta)u_0(k)+y_0(k)$}
        \STATE{$\qquad$ desired output:}
        \STATE{$\qquad\quad\tilde{y}(\theta,k)\leftarrow G_m(T_c,z)\tilde{r}(\theta,k)$}
        \STATE{$\qquad$ fictitious input:}
        \STATE{$\qquad\quad\tilde{u}(\theta,k)\leftarrow C(\theta,z)(\tilde{r}(\theta,k)-\tilde{y}(\theta,k))$}
        \STATE{\textbf{Step 3}: Minimize:}
        \STATE{$\qquad\quad J_{\mathrm{EFRIT}}(\theta)=\sum_{k=1}^{N}\left[y_0(k)-G_m(z)\tilde{r}(\theta,k)\right]^2$}
        \STATE{$\qquad\qquad\qquad\qquad+\lambda\sum_{k=1}^{N}\Delta\tilde{u}^2(\theta,k)$}
    \end{algorithmic}
\end{algorithm}

\subsection{A-FRIT}
\begin{figure}[t]
\centering
\includegraphics[width=8.5cm]{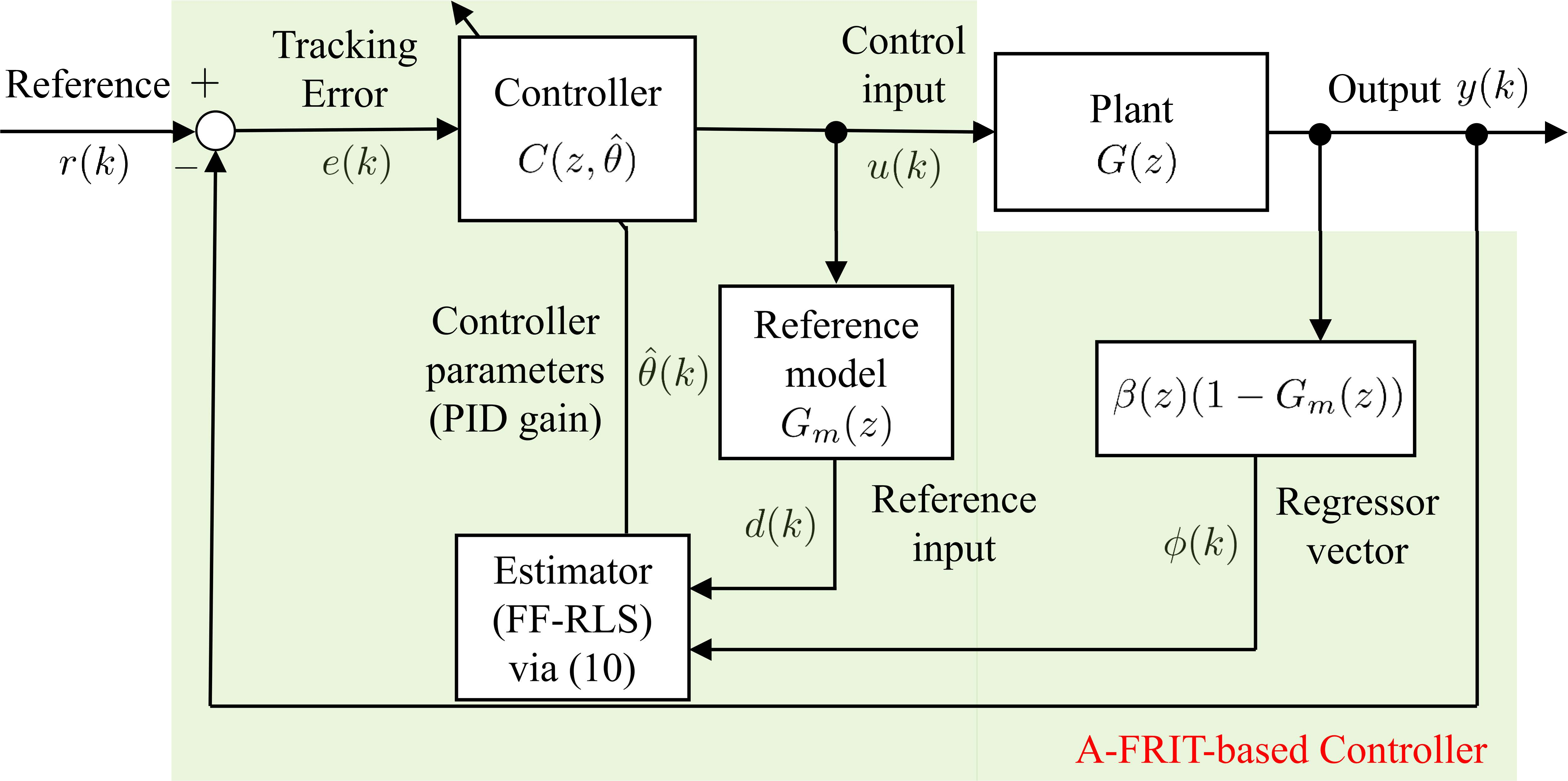}
  \caption{Block diagram of the adaptive FRIT method}
  \label{fig:A-FRIT}
\end{figure}

To update the controller parameters recursively, (\ref{ma:eva_FRIT}) must be transformed into a convex problem using the following lemma:
\begin{lem}
Assume that there exists a controller parameter $\theta$ such that
\begin{equation}\label{ma:ass}
y_0(k)-G_m(z)\tilde{r}(\theta,k)\equiv0,\ ^\forall k\geq 0.
\end{equation}
Then, by replacing the initial I/O data, $u_0(k)$ and $y_0(k)$ with I/O data, $u(k)$ and $y(k)$ for each time step, we obtain the convex evaluation function as follows:
\begin{equation}
\theta^{*} = \mathrm{arg}\ \min_\theta\ J_{\mathrm{A-FRIT}}(\theta),
\end{equation}
\begin{equation}\label{ma:eva_A-FRIT}
J_{\mathrm{A-FRIT}}(\theta)=\sum_{k=1}^{N}\hat{e}^2(k)\triangleq\sum_{k=1}^{N}\left[\phi^T(k)\theta-d(k)\right]^2,
\end{equation}
where regressor vector $\phi(k)\triangleq \beta(z)\{1-G_m(z)\}y(k)$ and the reference input $d(k)\triangleq G_m(z)u(k)$ are I/O data.
\end{lem}
\begin{pf}
Refer to \cite{A-FRIT1,A-FRIT2}.
\end{pf}
Using Lemma 1, the recursive least square (RLS) algorithm \cite{RLS1,RLS2} can be derived by minimizing (\ref{ma:eva_A-FRIT}), and adaptive PID control can be easily realized.
Its block diagram is shown in {\bf Fig. \ref{fig:A-FRIT}},
where $\hat{\theta}(k)$ is an estimate of $\theta(k)$.
\begin{algorithm}[t]
    \caption{A-FRIT algorithm}\label{alg1}
    \begin{algorithmic}
        \REQUIRE {$\beta(z)$, $G_m(z)$, $\hat{\theta}(0)$}
        \ENSURE {$u(k)$}
        \FORALL{$k\geq 0$}
        \STATE{$\phi(k)\leftarrow \beta(z)\{1-G_m(z)\}y(k)$}
        \STATE{$d(k)\leftarrow G_m(z)u(k)$}
        \STATE{$\hat{\theta}(k)\leftarrow\textrm{Estimation algorithms (e.g.\ Algorithm \ref{alg2})}$}
        \STATE{$u(k)\leftarrow \hat{\theta}^T(k)\beta(z)(r(k)-y(k))$}
        \ENDFOR
    \end{algorithmic}
\end{algorithm}

Next, we consider the update law for the estimated controller parameter $\hat{\theta}(k)$.
In general, we can apply the several adaptive algorithm, e.g. RLS, normalized gradient algorithm.
In this study, to improve robustness against time-varying systems and ensure the feasibility of the systems, we also describe an extension to A-FRIT using DF-based RLS that introduces the DF algorithm \cite{DF}, which is a type of forgetting factor.
A forgetting factor algorithm for RLS has been surveyed \cite{FF}, in particular, the commonly used exponential forgetting (EF) algorithm is effective for time-varying systems.
However, the EF algorithm cannot guarantee stability if the persistent excitation (PE) condition of the regressor vector (see Definition 1) is not satisfied.
By contrast, the DF algorithm is a more robust method because it can guarantee stability regardless of the PE condition by orthogonal decomposition of the information matrix $R(k)\in\mathbb{R}^{n\times n}$ comprising the regressor vector, which means the inverse of covariance matrix $P(k)\in\mathbb{R}^{n\times n}$.
To explain the effectiveness of DF algorithm, the decomposition of the information matrix for DF is expressed as follows:
\begin{equation}\label{ma:information_matrix}
R(k)=R_1(k-1)+\mu R_2(k-1)+\phi(k)\phi^T(k)
\end{equation}
where positive semidefinite matrix $R_1(k-1)\in\mathbb{R}^{n\times n}\ \mathrm{s.t.}\ \phi(k)\neq 0$ and its rank is $n-1$, and positive semidefinite matrix $R_2(k-1)\in\mathbb{R}^{n\times n}\ \mathrm{s.t.}\ R_2(k-1)\phi(k)=R(k-1)\phi(k)$ and its rank is $1$.
Also, $\mu\in (0,1]\in\mathbb{R}$ denotes the forgetting factor value.
This equation (\ref{ma:information_matrix}) forgets only in the direction of $\phi(k)\phi^T(k)$, i.e., by multiplying only $R_2(k-1)$ by $\mu$.
Then, the positive definiteness of the information matrix is always preserved.
Hence, DF-based A-FRIT is feasible without PE condition, and we can avoid estimation windup \cite{DF}.
Moreover, let $\varepsilon>0\in\mathbb{R}$ be the design parameter, which is determined by the signal-to-noise ratio and functions as a deadzone for the update.
The DF-based RLS algorithm for A-FRIT is expressed as Algorithm 3.

\begin{algorithm}[t]
    \caption{DF-based RLS algorithm for AFRIT}\label{alg2}
    \begin{algorithmic}
        \REQUIRE {$\mu\in(0,1)$, $\theta(0)$, positive-definite $P(0)$, $\varepsilon >0$}
        \ENSURE {$\hat{\theta}(k)$}
        \FORALL{$k\geq 0$}
        \IF{$\|\phi(k)\|>\varepsilon$}
        \STATE {$\bar{P}(k-1)\leftarrow P(k-1)+\frac{1-\mu}{\mu}\frac{\phi(k)\phi^T(k)}{\phi^T(k)R(k-1)\phi(k)}$}
        \STATE{$M(k)\leftarrow (1-\mu)\frac{R(k-1)\phi(k)\phi^T(k)}{\phi^T(k)R(k-1)\phi(k)}$}
        \ELSE
        \STATE {$\bar{P}(k-1)\leftarrow P(k-1)$}
        \STATE {$M(k)\leftarrow 0$}
        \ENDIF
        \STATE{$R(k)\leftarrow [I-M(k)]R(k-1)+\phi(k)\phi^T(k)$}
        \STATE{$P(k)\leftarrow\bar{P}(k-1)-\frac{\bar{P}(k-1)\phi(k)\phi^T(k)\bar{P}(k-1)}{\phi^T(k)\bar{P}(k-1)\phi(k)}$}
        \STATE{$\hat{\theta}(k)\leftarrow\hat{\theta}(k-1)+P(k)\phi(k)[d(k)-\phi^T(k)\hat{\theta}(k-1)]$}
        \ENDFOR
    \end{algorithmic}
\end{algorithm}

\begin{rem}
A-FRIT can be realized using various forgetting factors.
In \cite{A-FRIT2}, several forgetting factors were experimentally compared in detail using the A-FRIT algorithm.
Refer to the aforementioned paper for further details.
\end{rem}

\begin{rem}
The PE condition for a regressor vector is represented by the following definition.
This condition indicates that there always exists an interval in which the minimum eigenvalue of the matrix consisting regressor vectors is positive.
However, it is known that satisfying this condition is not realistic in adaptive control because the target trajectory is arbitrarily determined.
\begin{dfn}\cite{Adaptive}
A bounded regressor vector $\phi(k)\in\mathbb{R}^n,\ n\geq 1$, is persistently exciting (PE) if there exists $\delta>0$ and $\alpha_0>0$ such that
\begin{equation}
\sum_{k=\sigma}^{\sigma+\delta}\phi(k)\phi^T(k)\geq \alpha_0 I,\ ^\forall\sigma\geq 0
\end{equation}
\end{dfn}
\end{rem}

\section{Proposed method}
In this section, we describe the PL model utilizing adaptive model matching of the A-FRIT algorithm and the model predictive control system based on this model.

\subsection{PL model utilizing A-FRIT}
FRIT, E-FRIT and A-FRIT algorithms tune controller parameters such that the closed-loop system matches the reference model with ideal characteristics.
However, designing this reference model is difficult for data-driven control methods that assume that the characteristics of the plant are unknown.
If an inappropriate reference model is selected, control performance can be significantly degraded.
Therefore, we focus on the model-matching problem, which is the basic concept of FRIT, and separate it from the trajectory-tracking problem.
Using a linear model as the model to be matched, the closed-loop system can be considered a linear system, which implies the PL technique, this is known as the PL model.
In this study, we improved the model matching accuracy using an adaptive model matching based on A-FRIT.
Specifically, we set the PL model to the following state-space model:
\begin{equation}\label{ma:PL}
\left\{
\begin{split}
x(k+1) &= a_px(k)+b_pu_c(k)\\
y(k) &= c_px(k)
\end{split}
\right.,
\end{equation}
with $a_p= \exp(-{T_s}/{T_c})$, $b_p= 1-\exp(-{T_s}/{T_c})$, and $c_p=1$,
where $T_c$ denotes the time constant value.
In addition, $x(k)$, $u_c(k)$, and $y(k)\in\mathbb{R}$ denote the state variable, optimal control input, and output, respectively.
Note that $u_c(k)$ is a virtual target trajectory from the perspective of the PL model.

\begin{figure}[t]
\centering
\includegraphics[width=8.5cm]{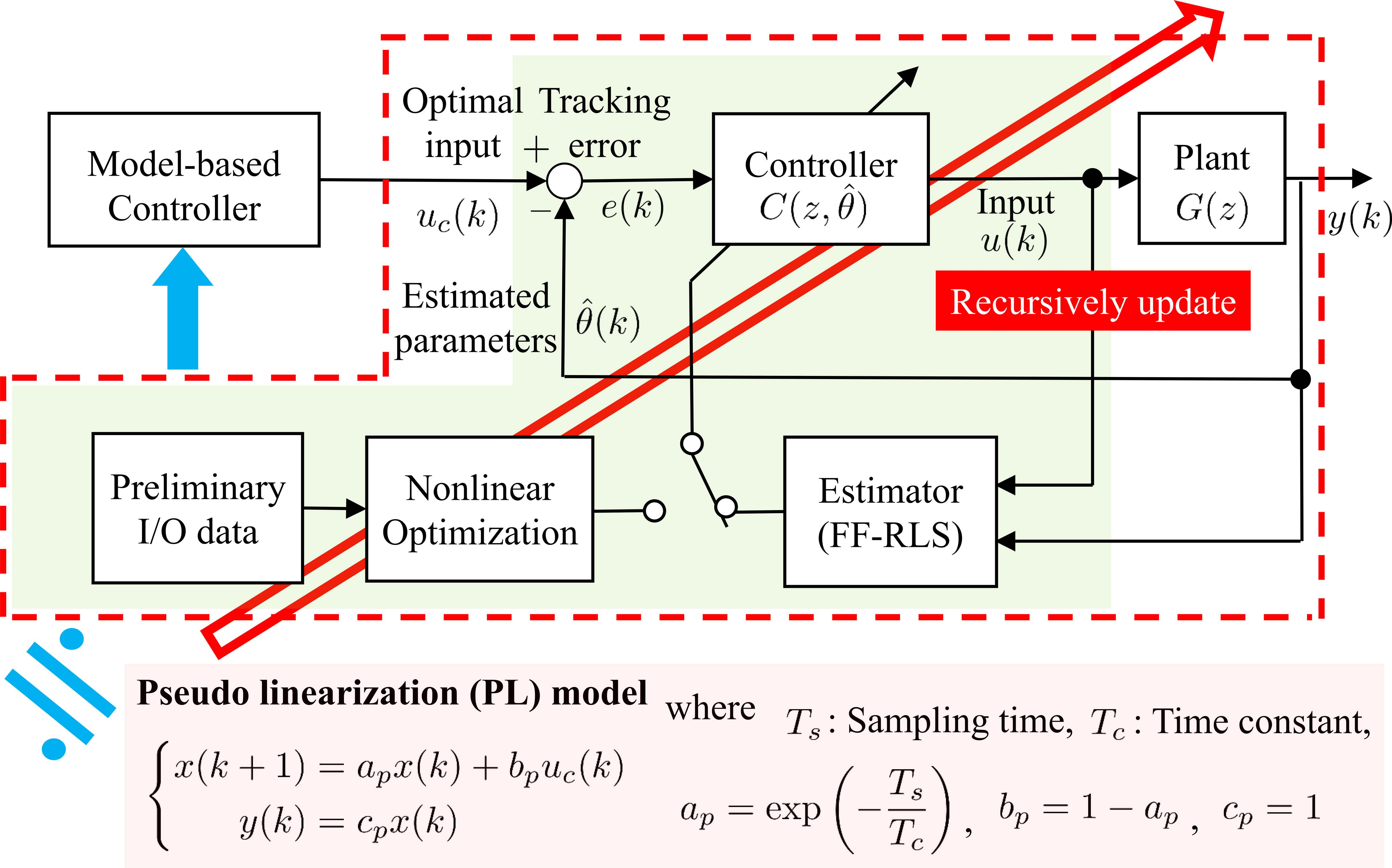}
  \caption{Conceptual diagram of direct data-driven adaptive model matching}
  \label{fig:DDAMM}
\end{figure}

\begin{figure}[t]
\centering
\includegraphics[width=8.5cm]{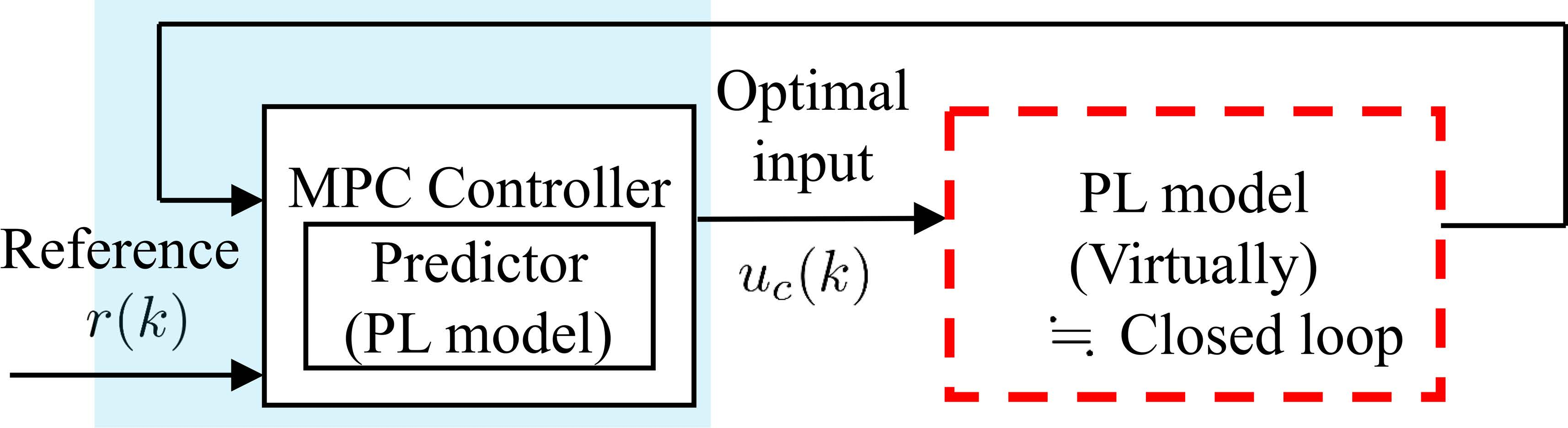}
  \caption{Conceptual diagram of MPC design using PL model}
  \label{fig:PLMPC}
\end{figure}

{\bf Figures \ref{fig:DDAMM}} and {\bf \ref{fig:PLMPC}} summarize concepts above.
From {\bf Fig. \ref{fig:DDAMM}}, the PID gain as the initial gain is obtained based on E-FRIT through preliminary experiments.
Then, data-driven adaptive model matching estimates the optimal PID gains to match the PL model using only measured I/O data at each time step.
Hence, the entire inner loop is updated recursively, enabling model matching at all steps, independent of prior experimental data.
The model-based control system is then designed in the outer loop based on the PL model.
In this paper, MPC is designed based on the PL model as shown in {\bf Fig. \ref{fig:PLMPC}}.
In this case, the inner loop becomes a virtual PL model, thus, it can predict appropriately, and input constraints can be considered.
The design of the external loop is described in Section 3.2.

Next, we discuss the design parameter $T_c$.
In our previous study, the existence of an optimal design value, $T_c$ \cite{FMPC1} was illustrated.
In FRIT, as the smaller the value of $T_c$, the evaluation function shown in (\ref{ma:eva_FRIT}) is minimized, because the response of the reference model approaches the target trajectory.
On the other hand, in E-FRIT, a smaller time constant $T_c$ does not necessarily result in a smaller evaluation function, since the change term of the input is included in the evaluation function (\ref{ma:eva_EFRIT}).
Therefore, the control performance of the E-FRIT strongly depends on the time constant $T_c$ of the reference model, however determining this value is not easy due to the unknown plant.
Therefore, we have introduced a method \cite{E-FRIT2} in which the time constant of the reference model is also optimized simultaneously with the controller parameters.
In this study, $T_c$ is also included in the controller parameter set of E-FRIT as $\bar{\theta}=[\theta^T,\ T_c]^T=[K_p,\ K_i,\ K_d,\ T_c]^T$ for optimization, which also eliminates the requirement for trial and error in the PL model.
Thus, in Algorithm 1, $\theta$ is replaced with $\bar{\theta}$ and $G_m (z)$ with the PL model.
Note that, in the case of the A-FRIT algorithm, it can be updated at each time without any prior optimization; however, it is preferable to optimize it with E-FRIT in advance to improve transient performance.
\begin{rem}
The difference from the conventional reference model is that the PL model has no physical interpretation other than the output generator.
Therefore, the PL model should have only one state variable, and a model with the characteristics of a first-order system was employed.
\end{rem}

\begin{rem}
As the time constant of the reference model is significantly reduced, FRIT and A-FRIT loses control performance.
This is because the time constant cannot be considered in the evaluation function, and if the time constant is too small, controller gain becomes very high, and large overshoots occur.
The proposed method solves these problems because it considers input constraints.
\end{rem}

\begin{rem}
Model-based adaptive model matching \cite{MBAMM} requires first applying feedback to linearize the system and after that designing feedback and feedforward compensators, respectively.
On the other hand, the proposed method achieves model matching to a PL model using only a PID controller via A-FRIT.
Therefore, it is a novel data-driven adaptive model matching method without requiring a mathematical model.
\end{rem}

\subsection{Adaptive model predictive controller design}
\begin{figure}[t]
\centering
\includegraphics[width=8.5cm]{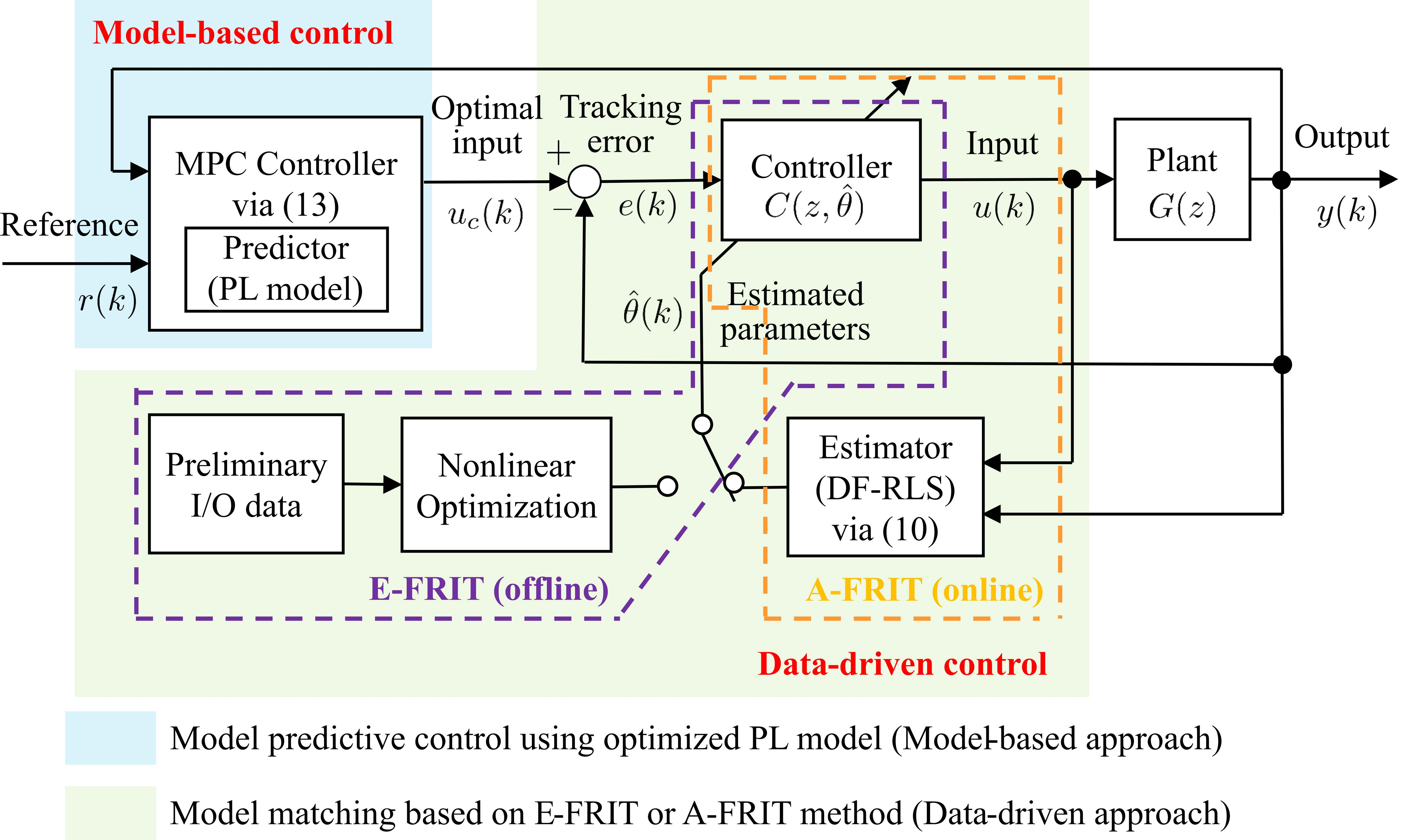}
  \caption{Block diagram of control method; FRIT-based MPC (FMPC) and adaptive FRIT-based MPC (A-FMPC)}
  \label{fig:A-FRIT-MPC}
\end{figure}
In the inner loop, the closed-loop system is matched to the PL model using the adaptive model matching based on A-FRIT, and the MPC is designed in the outer loop, as shown in Fig. \ref{fig:A-FRIT-MPC}.
Hence, the PL model was treated as a predictor of MPC, and the control system was designed accordingly.
The evaluation function of MPC is expressed as 
\begin{equation}\label{ma:eva_MPCS}
\begin{split}
&J(k)=J_e(k)+J_{u_c}(k)+J_u(k)\\
&\qquad\mathrm{subject\ to}\ u_{\min}\leq \hat{u}(k+i)\leq u_{\max},\ i=0,\ldots, H_u-1,
\end{split}
\end{equation}
where $J_e(k)$, $J_{u_c}(k)$, and $J_u(k)$ denote respectively the evaluation function for error, optimal input difference, and actual input difference using each weights $Q(i)\geq 0$, $R(i)> 0$, and $R_{u}(i)\geq 0$ as follows:
\begin{equation}
\left\{
\begin{split}
J_e(k)&=\sum_{i=1}^{H_p}\left\|\hat{y}(k+i|k)-r(k+i|k)\right\|^2_{Q(i)},\\
J_{u_c}(k)&=\sum_{i=0}^{H_u-1}\left\|\Delta u_c(k+i|k)\right\|^2_{R(i)},\\
J_u(k)&=\sum_{i=0}^{H_u-1}\left\|\Delta \hat{u}(k+i|k)\right\|^2_{R_u(i)},\\
\end{split}
\right.
\end{equation}
$\hat{y}(k)$ and $\hat{u}(k)$ denote the estimated value of $y(k)$ and $u(k)$ using the PL model, respectively.
Moreover, each variable is obtained as $\Delta u_c(k)\triangleq u_c(k)-u_c(k-1)$, $\Delta \hat{u}(k)\triangleq \hat{u}(k)-\hat{u}(k-1)$.
In addition, $H_p\in\mathbb{Z^+}$ and $H_u\in\mathbb{Z^+}$ denote the prediction and control horizons, respectively.

\begin{rem}
The evaluation term $J_u(k)$ is introduced when the control input $u(k)$ for plant exhibits large variations in applied inputs $u(k)$.
For a discussion of the effects of this evaluation term, refer to \cite{FMPC2}.
\end{rem}

Furthermore, to realize the aforementioned evaluation function and input constraints, an $i$-step-ahead prediction of the input is required.
Let $\hat{e}(k+i)\triangleq u_c(k+i)-\hat{y}(k+i),\ i=0,\ldots, H_u-1$ be the difference between the optimal input and predicted output, which is the error from the perspective of the PID controller.
We assume that the PID gain estimated by A-FRIT is maintained in the control horizon $H_u$.
The control input can then be predicted using the PL model as follows:
\begin{equation}
\hat{u}(k+i) = \hat{u}_p(k+i) + \hat{u}_i(k+i) + \hat{u}_d(k+i), i=0,\ldots,H_u-1,
\end{equation}
where
\begin{equation}
\left\{
\begin{split}
\hat{u}_p(k+i)&=\hat{K}_p(k)\hat{e}(k+i),\\
\hat{u}_i(k+i)&=\hat{u}_i(k+i-1)+\hat{K}_i(k)T_s\hat{e}(k+i),\\
\hat{u}_d(k+i)&=\hat{K}_d(k)\frac{\hat{e}(k+i)-\hat{e}(k+i-1)}{T_s}.\\
\end{split}
\right.
\end{equation}

\section{Experimental results and discussion}
To evaluate effectiveness of the proposed methods, we compare the three types of controller; 1) A-FRIT, 2) FMPC, and 3) A-FMPC.
Next, we evaluate the effect of the design parameter $\lambda$ in E-FRIT, and discuss the robustness against the proposed method.

\subsection{Experimental setup}
\begin{figure}[t]
\begin{minipage}[b]{0.49\linewidth}
\centering
\includegraphics[width=4.5cm]{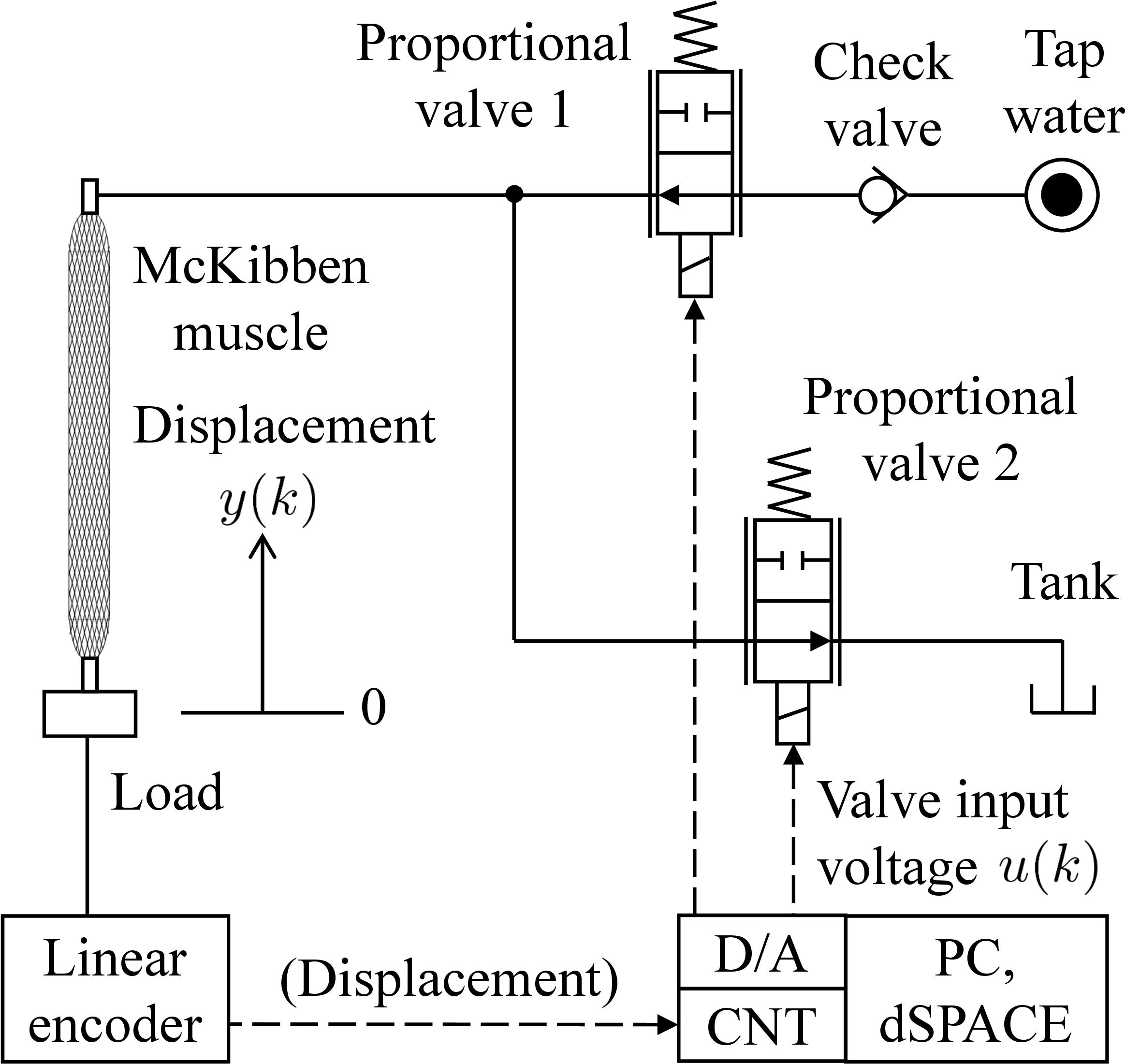}
  \subcaption{experimental circuit}
  \end{minipage}
 \begin{minipage}[b]{0.49\linewidth}
 \centering
 \includegraphics[width=3cm]{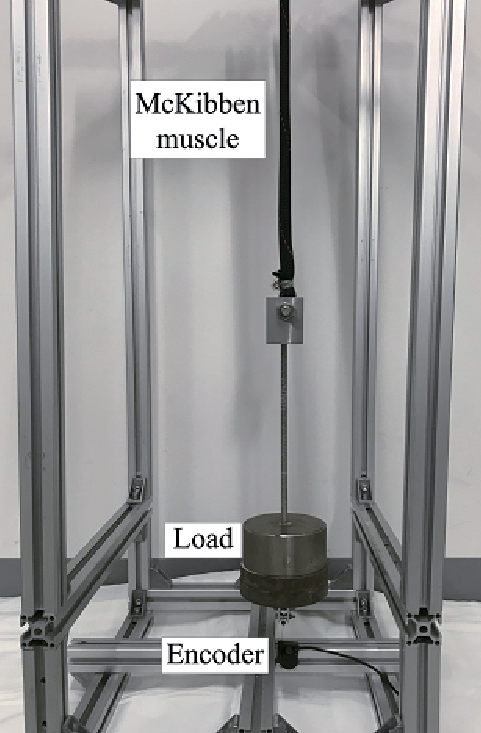}
 \subcaption{experimental equipment}
 \end{minipage}
 \caption{Experimental setup for water hydraulic artificial muscle:}
  \label{fig:setup}
\end{figure}

{\bf Figure \ref{fig:setup}} shows the experimental circuit and equipment, respectively.
The setup consists of a McKibben-type artificial muscle, two proportional valves, a linear encoder, and a controller PC.
The muscle contracts for increase of tap water pressure.
The proportional valve is a type of flow control, and its flow rate is controlled by the input voltage $u(k)$. The linear encoder measures the displacement, $y(k)$, of the muscle and sends it to the controller.
These details are summarized in Table \ref{table:spec}.

\begin{table}[h]
	\begin{center}
		\caption{Specifications of experimental components}
		\label{table:spec}
		\begin{tabular}{ll}\hline
		\rule[0mm]{0mm}{4mm}Item & Specifications\rule[0mm]{0mm}{4mm}\\\hline\hline 
		\rule[0mm]{0mm}{4mm}Proportional valves & KFPV300-2-80,\\
		& Koganei Corporation. \\
		& $C_v$ Value: 1.6;\\
		& Range of input voltage: 0 to 10 V\\
        & \\
        Pressure sensor & FP101,\\
        & Yokogawa Electric Corporation\\
        & Range: 0 to 1 MPa (abs)\\
		& \\
		Linear encoder & DX-025, MUTOH Industries Ltd. \\
		& Resolution: 0.01 mm\\
		& \\
		Controller PC & Operating system: Windows 10,\\
		& Microsoft Corporation. \\
		& CPU: 2.50 GHz,\\
		& RAM: 16.00 GB\\
		& Applications: MATLAB/Simulink\\
		& and dSPACE 1103\\
		& \\
		Tap-water-driven & Hand-made muscle,\\ muscle & Length: 400 mm\\
		& \\
		Tap-water & Average supply pressure:\\
		& 0.15 MPa (G)\\\hline
		\end{tabular}
	\end{center}
\end{table}

\subsection{Experimental conditions}
To evaluate the robustness against the proposed method and analyze the influence of optimized PID gain or initial PID gain for adaptive system using E-FRIT, we examine the three items; 1) changing the design parameter $\lambda$, 2) selecting the initial PID gain for pre-experiments, and 3) selecting the target trajectory for pre-experiments, i.e. the target trajectory of the pre-experiment and experiment are the same or different.

First, the PID gains for the inner-loop system were tuned using E-FRIT (see Algorithm 1).
In this study, the initial PID gains for E-FRIT were set to 
Case 1: $\theta_0 = [0.1,\ 0.1,\ 0.01]^T$, Case 2: $\theta_0 = [0.1,\ 0.1,\ 0.001]^T$, respectively.
In addition, the initial reference model for E-FRIT was set as $0.00995/(1-0.99z^{-1})$, which indicates a discrete-time first order transfer function with a time constant $T_c$ of $0.01$ s and a gain of $1$.
The design parameter $\lambda$ was set in nine values of $1.0$, $2.5$, $5.0$, $10$, $50$, $100$, $250$, $500$, and $1000$.
The initial I/O data were obtained for a closed-loop system for two target trajectories: a staircase square and a sinusoidal signal with an amplitude of $20$ mm, offset of $30$ mm, and frequency of $0.3$ Hz.
Based on the above conditions, E-FRIT optimized both the PID gains and the time constant values are listed in Tables \ref{table:case1} and \ref{table:case2}.
In addition, Table \ref{table:summary} shows the corresponding figures for experimental results and experimental conditions.
The obtained optimal PID gain $\theta^{*}$ is treated as a fixed gain in FMPC and as $\hat{\theta}(0)=\theta^{*}$ in A-FRIT or A-FMPC.

In the MPC design, the prediction and control horizons were set as $H_p=H_u=5$.
The weight matrices were set as $Q=I$, $R=40I$, and $R_u=I$.
In the design of the adaptive system, the initial values of the covariance matrix and information matrix were set to $P(0)=10^3I$ and $R(0)=10^{-3}I$, respectively.
Moreover, the threshold value and the forgetting factor for DF were set to $\varepsilon=10^{-3}$ and $\mu=0.99$.

\subsection{Experimental results}
\begin{figure}[h]
\centering
\includegraphics[width=8.5cm]{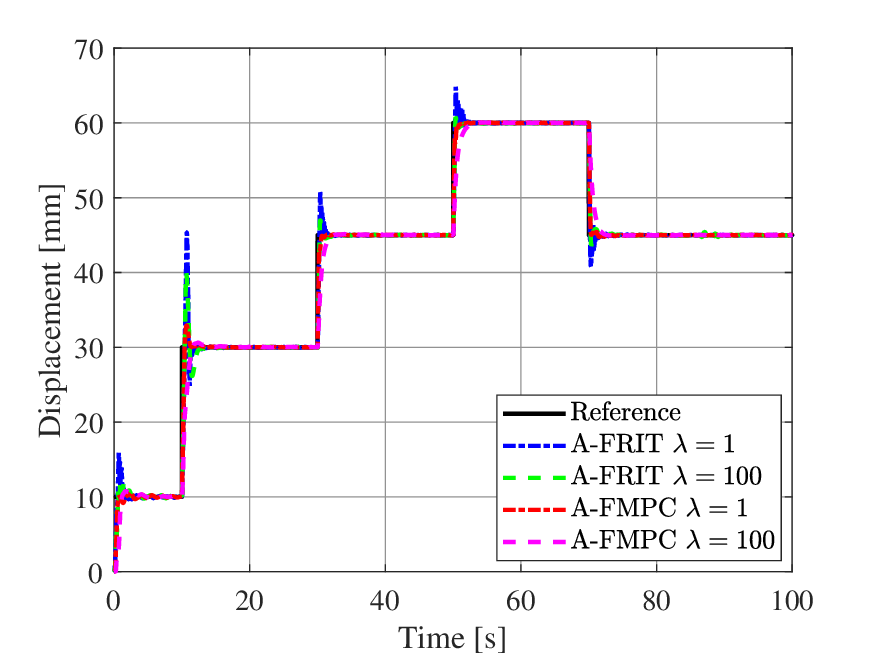}
  \caption{Comparison of control performance between A-FRIT (conv.) and A-FMPC (prop.) for the design parameter in E-FRIT $\lambda=1,100$}
  \label{fig:y_A-FRIT}
\end{figure}

\begin{figure}[h]
\centering
\includegraphics[width=8.5cm]{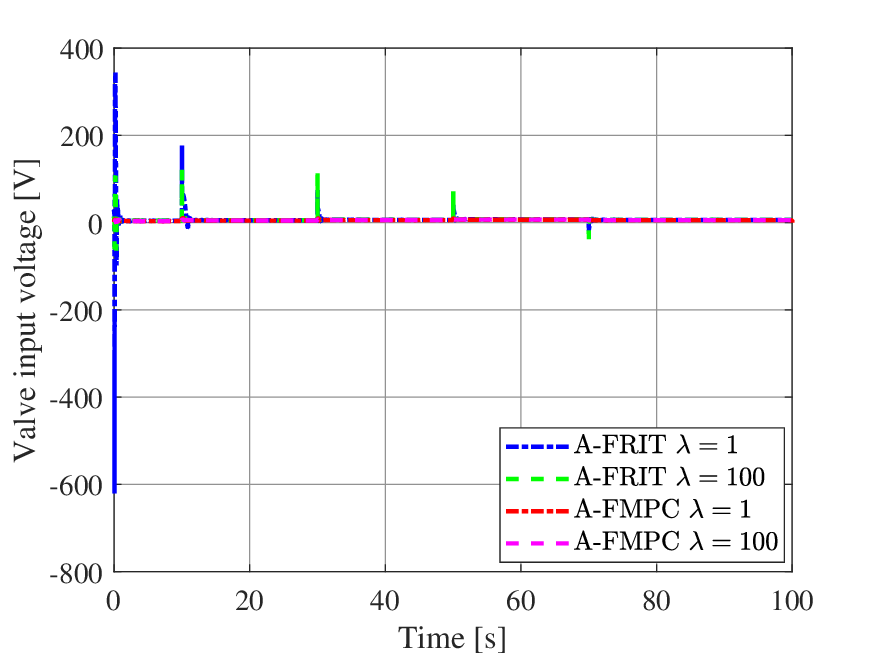}
  \caption{Comparison of control input between A-FRIT (conv.) and A-FMPC (prop.) for the design parameter in E-FRIT $\lambda=1,100$}
  \label{fig:u_A-FRIT}
\end{figure}

\begin{figure}[h]
\centering
\includegraphics[width=8.5cm]{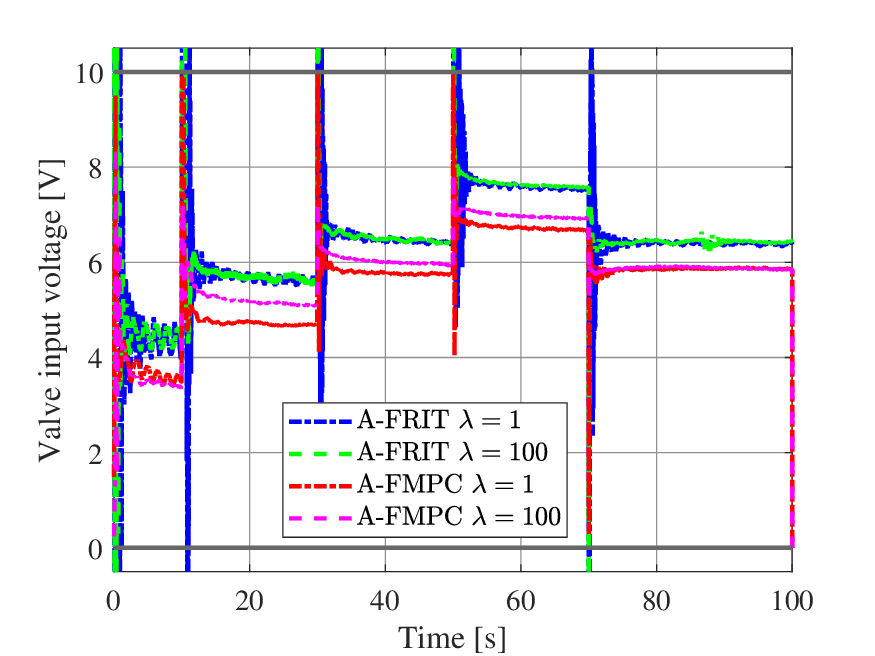}
  \caption{Enlarged view of Fig. \ref{fig:u_A-FRIT} with input constrainsts (Black line: 0-10 V)}
  \label{fig:u_A-FRIT2}
\end{figure}

\begin{figure}[h]
\centering
\includegraphics[width=8.5cm]{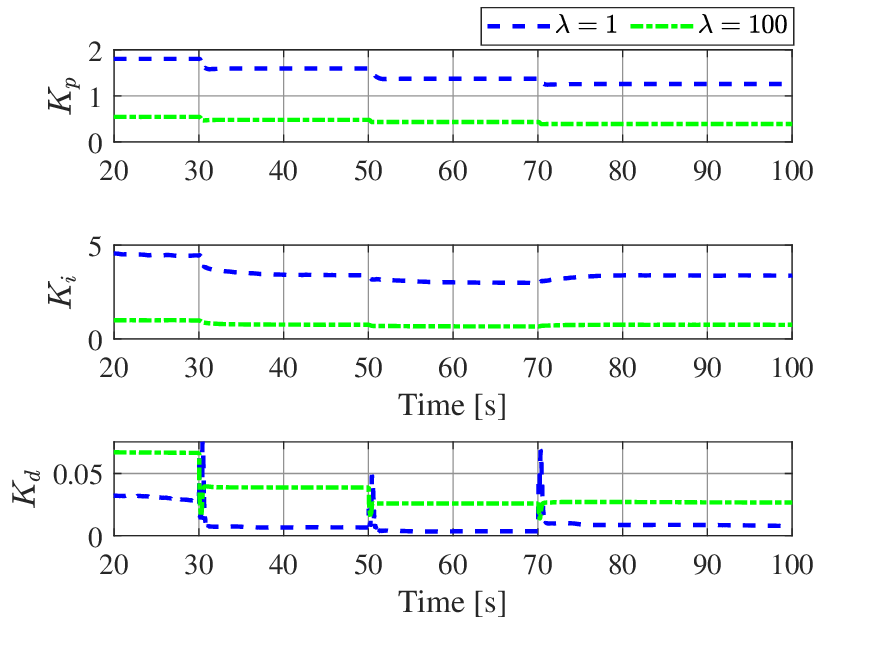}
  \caption{Comparison of estimated PID gains for A-FRIT in the steady-state response (20-100 s) by the design parameter $\lambda=1,100$}
  \label{fig:pid_A-FRIT}
\end{figure}

First, we compared the conventional method, A-FRIT, with the proposed method, A-FMPC.
To evaluate the influence of the time constant of the optimized PL model, {\bf Figs. \ref{fig:y_A-FRIT}}, {\bf \ref{fig:u_A-FRIT}} and {\bf\ref{fig:u_A-FRIT2}} show that the control performance in the initial gain and the optimal time constant for two cases, $\lambda=1$, $\lambda=100$, are discussed.
From {\bf Fig. \ref{fig:y_A-FRIT}}, A-FRIT has a larger overshoot when the design parameter $\lambda=1$ than when $\lambda=100$.
This is because a smaller value of $\lambda$ produces high PI gain as shown in {\bf Fig. \ref{fig:pid_A-FRIT}}, thus higher control input shown in {\bf Figs. \ref{fig:u_A-FRIT}} and {\bf \ref{fig:u_A-FRIT2}}.
On the other hand, since the A-FMPC satisfies the input constraints as shown in {\bf Figs. \ref{fig:u_A-FRIT}} and {\bf \ref{fig:u_A-FRIT2}}, it does not generate excessive control inputs nor overshoot, resulting in a significant improvement in transient response.
It also means that the control system relaxed strong dependency of the design parameter $\lambda$.

\begin{figure}[h]
\centering
\includegraphics[width=8.5cm]{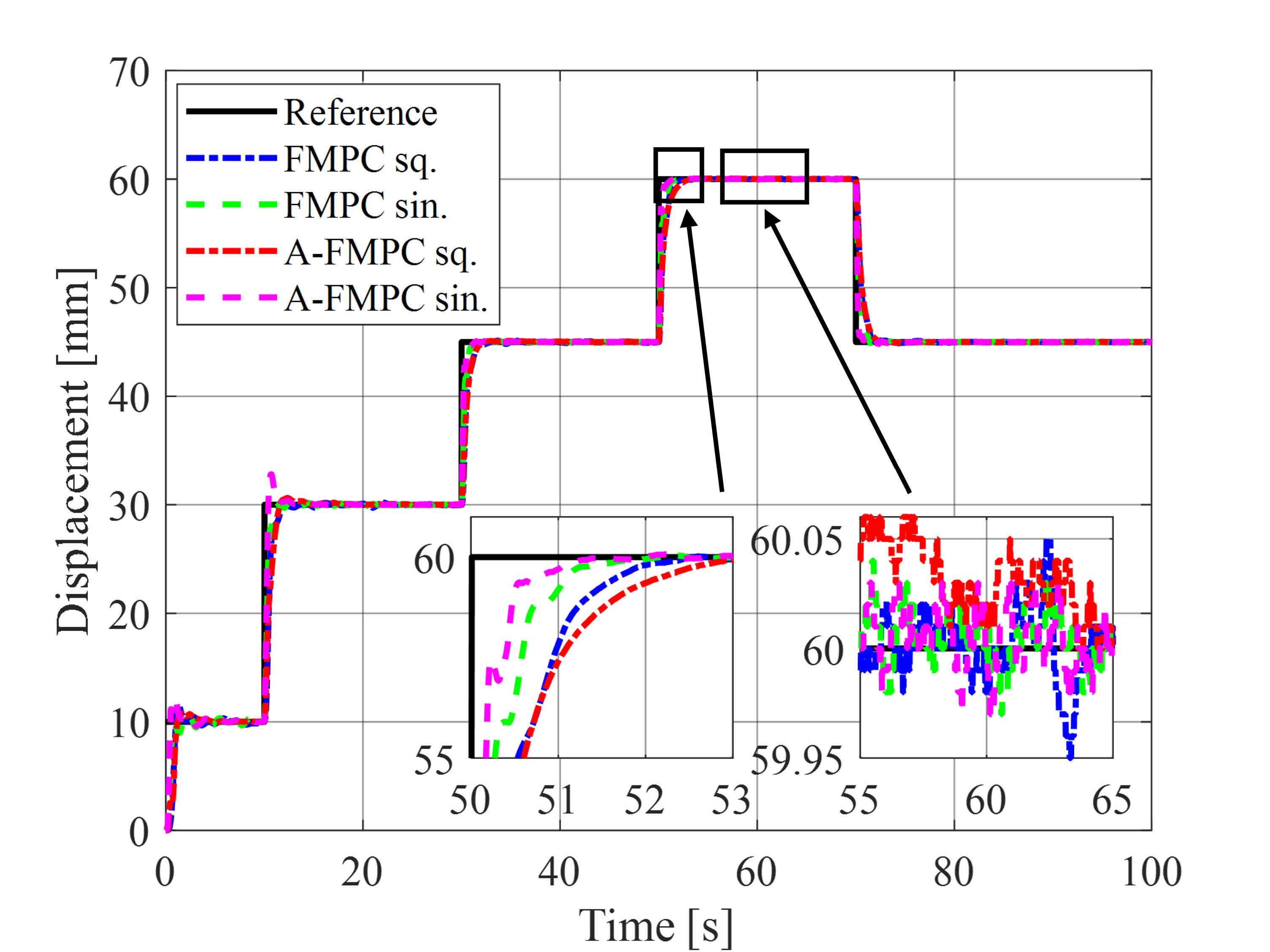}
  \caption{Comparison of control performances between FMPC (conv.) and A-FMPC (prop.) for parameter tuning by E-FRIT with two different target trajectories (sinusoidal or square signal) under Case 1 (initial PID gain $\theta=[0.1,\ 0.1,\ 0.01]^T$)}
  \label{fig:y_FMPC}
\end{figure}

\begin{figure}[h]
\centering
\includegraphics[width=8.5cm]{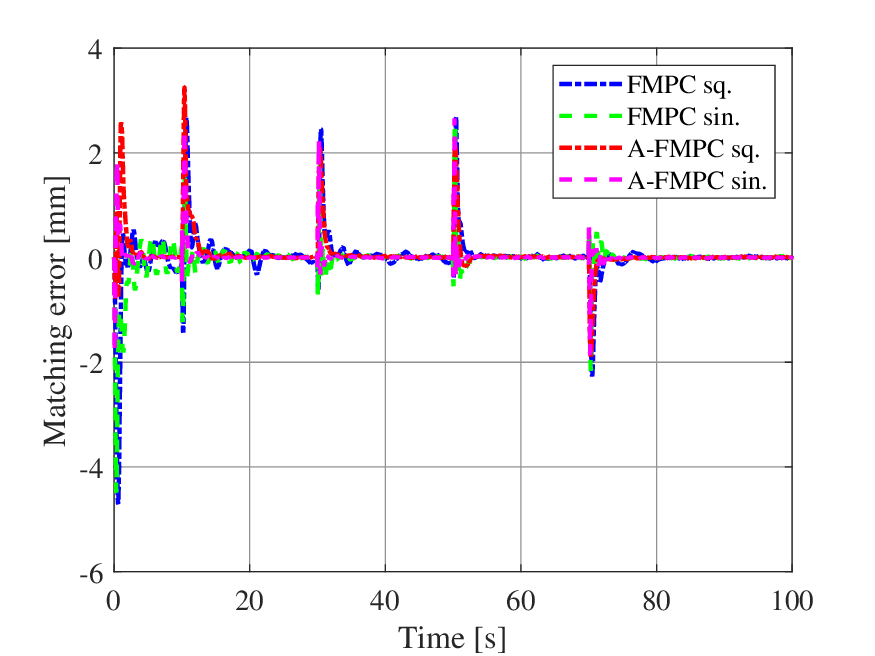}
  \caption{Comparison of matching errors, which are the errors between the PL model output and the measured displacement of the muscle, between FMPC (conv.) and A-FMPC (prop.) for parameter tuning by E-FRIT with two different target trajectories (sinusoidal or square signal) under Case 1 (initial PID gain $\theta=[0.1,\ 0.1,\ 0.01]^T$)}
  \label{fig:me_FMPC}
\end{figure}

\begin{figure}[h]
\centering
\includegraphics[width=8.5cm]{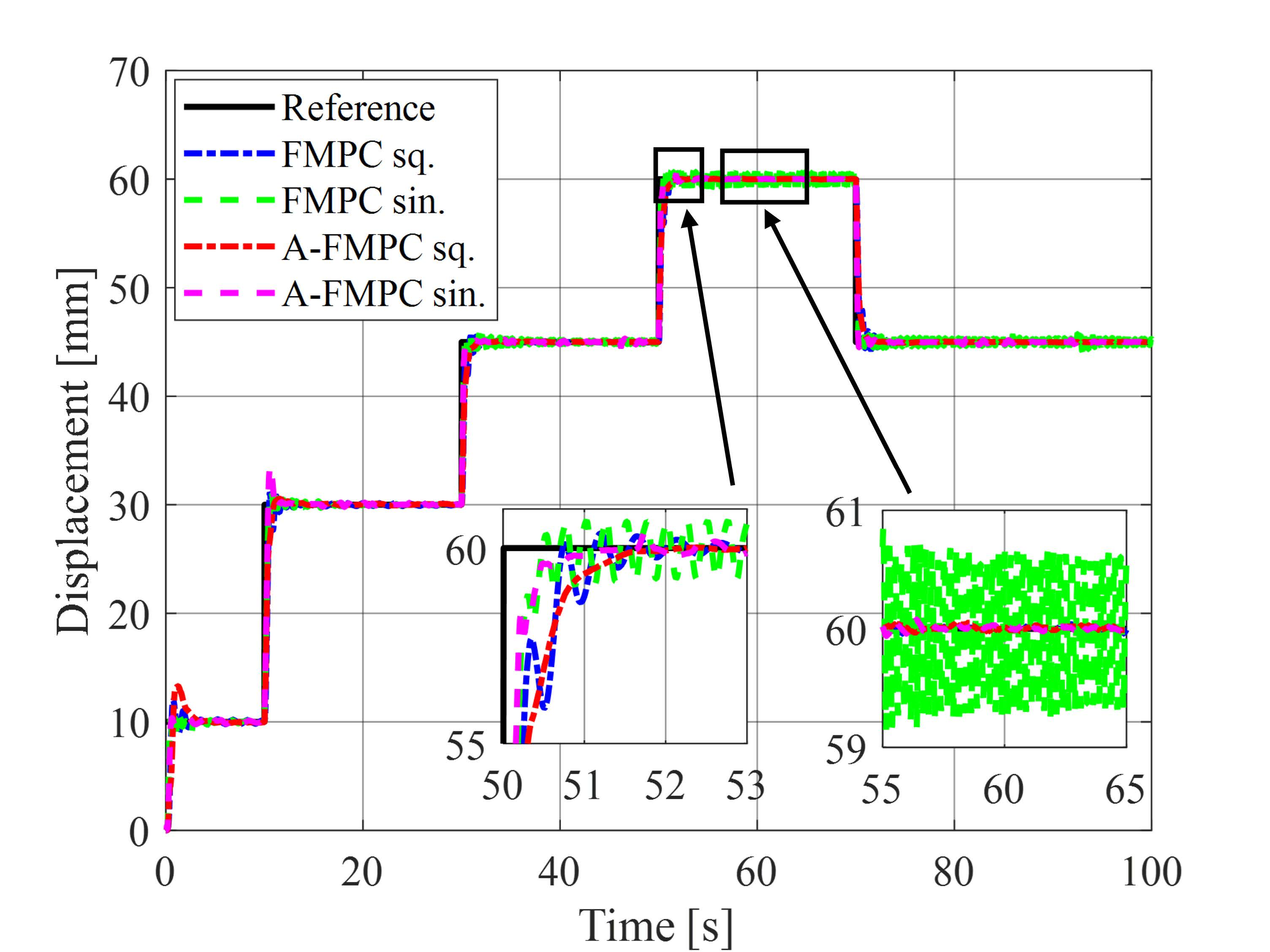}
  \caption{Comparison of control performances between FMPC (conv.) and A-FMPC (prop.) for parameter tuning by E-FRIT with two different target trajectories (sinusoidal or square signal) under Case 2 (initial PID gain $\theta=[0.1,\ 0.1,\ 0.001]^T$)}
  \label{fig:y_FMPC2}
\end{figure}

\begin{figure}[h]
\centering
\includegraphics[width=8.5cm]{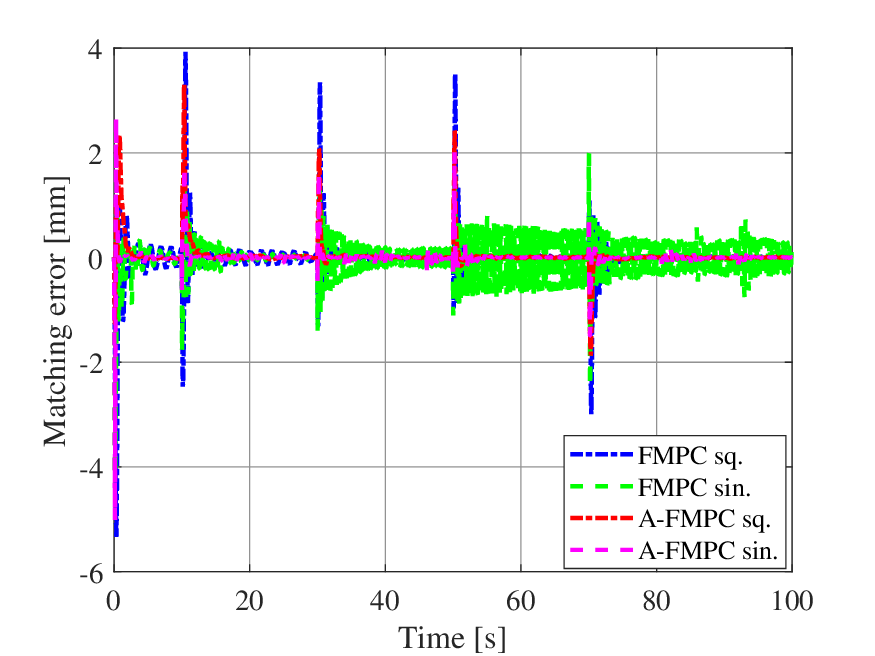}
  \caption{Comparison of matching errors, which are the errors between the PL model output and the measured displacement of the muscle, between FMPC (conv.) and A-FMPC (prop.) for parameter tuning by E-FRIT with two different target trajectories (sinusoidal or square signal) under Case 2 (initial PID gain $\theta=[0.1,\ 0.1,\ 0.001]^T$)}
  \label{fig:me_FMPC2}
\end{figure}

Next, we discuss the comparison of FMPC and A-FMPC between the case with a square signal and the case with a sinusoidal signal in the preliminary experiments for parameter tuning.
In this case, the initial PID gain uses Case 1 and 2, and the design parameter $\lambda$ was set to $100$.
{\bf Figure \ref{fig:y_FMPC}} compares the control performance in Case 1.
From the figure, the control performance of the four conditions was almost equivalent.
Hence, in this case, the effect of changing the optimal time constant of the PL model is small, and the impact of different target trajectories during parameter tuning and control is also small for both the FMPC and A-FMPC methods.
The matching error, which is the difference between the displacement and the output of the PL model shown in {\bf Fig. \ref{fig:me_FMPC}}, is also reduced for all cases, thus it is reasonable.
Therefore, the control performance appears to be independent of the prior experimental data.
On the other hand, {\bf Fig. \ref{fig:y_FMPC2}} compares the control performance in Case 2.
This figure shows that the control performance of FMPC does not deteriorate when the same target trajectory is used during tuning and control, but when the target trajectory is different, the response becomes very oscillatory and does not track the target trajectory.
The matching error shown in {\bf Fig. \ref{fig:me_FMPC2}} is not reduced, indicating that model matching is not properly performed.
This is because the FRIT/E-FRIT are tuned by nonlinear optimization, and the optimization result strongly depends on the initial PID gain value, i.e., initial value dependence.
In this case, from Table 3, E-FRIT for Case 2 generates a very small value of D-gain due to setting it to 1/10th of the value in Case 1, which in turn generates a higher PI-gain.
In addition, the optimized time constant is smaller, which corresponds to the higher PI gain generated.
Selecting the initial PID gain is very important because control performance depends strongly on the initial PID gain.
However, determining the initial PID gain by trial and error is contrary to the main purpose of this study, and leads to a significant the time-consuming because the design method for the initial PID gain has not been systematic.
On the other hand, the proposed, A-FMPC method, shows no degradation in control performance in either Case 1 or 2.
This is because FMPC uses a fixed PID gain, while A-FMPC adaptively updates the PID gain at each time step.
Hence, the proposed method is independent of prior experimental data and can achieve high robustness.

\begin{figure}[t]
\centering
\includegraphics[width=8.5cm]{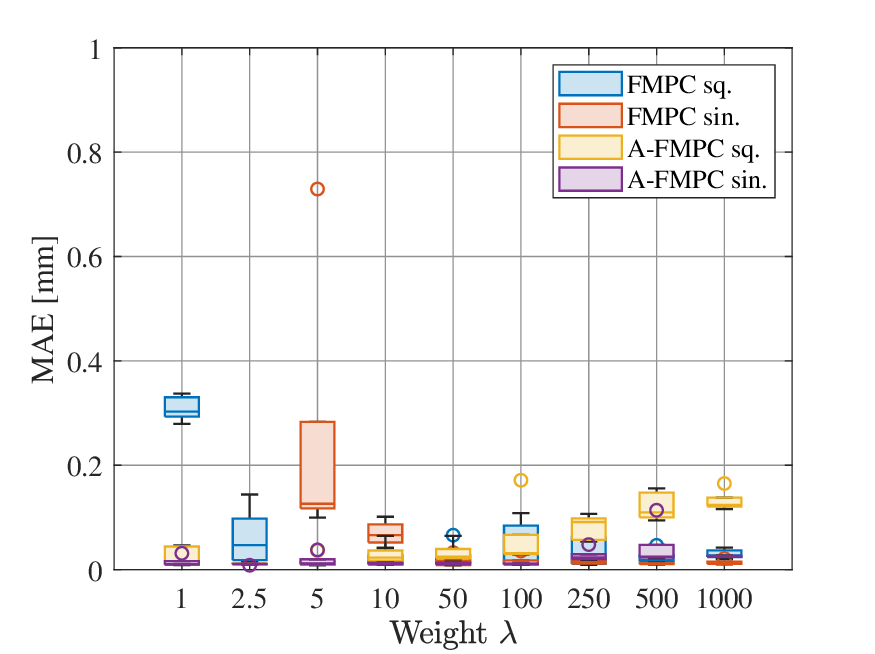}
  \caption{Comparison of box plots for changing the design parameter $\lambda$ between FMPC (conv.) and A-FMPC (prop.) for parameter tuning by E-FRIT with two different target trajectories (sinusoidal or square signal) under Case 1 (initial PID gain $\theta=[0.1,\ 0.1,\ 0.01]^T$) in the steady state response (55-65 s)}
  \label{fig:boxplot_ste}
\end{figure}

\begin{figure}[t]
\centering
\includegraphics[width=8.5cm]{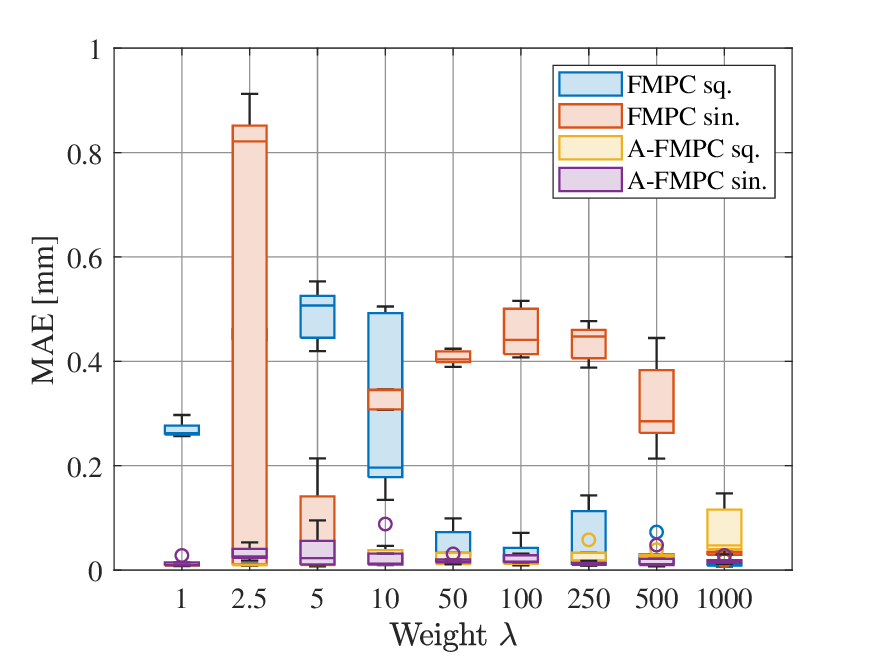}
  \caption{Comparison of box plots for changing the design parameter $\lambda$ between FMPC (conv.) and A-FMPC (prop.) for parameter tuning by E-FRIT with two different target trajectories (sinusoidal or square signal) under Case 2 (initial PID gain $\theta=[0.1,\ 0.1,\ 0.001]^T$) in the steady state response (55-65 s)}
  \label{fig:boxplot_ste2}
\end{figure}

\begin{figure}[t]
\centering
\includegraphics[width=8.5cm]{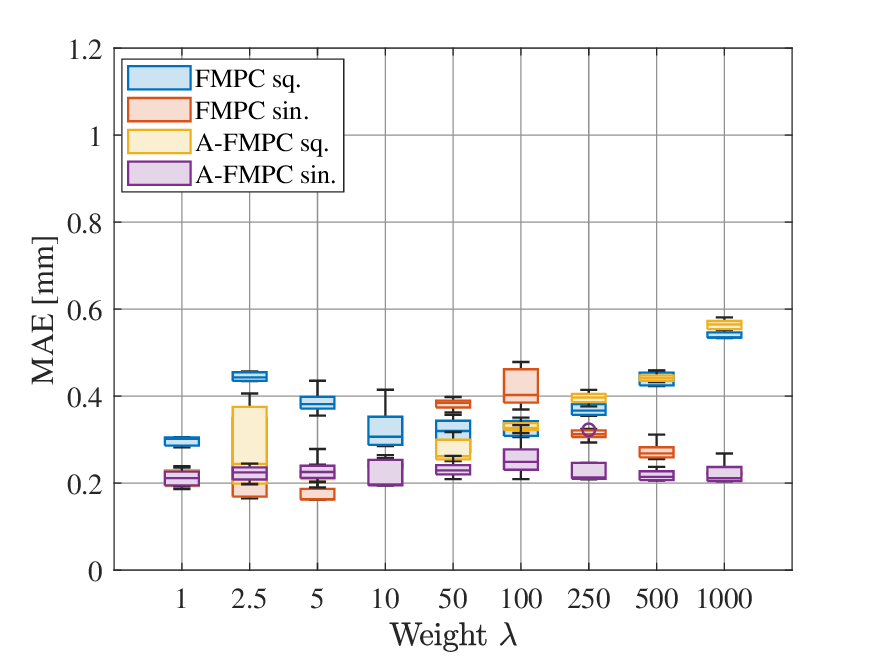}
  \caption{Comparison of box plots for changing the design parameter $\lambda$ between FMPC (conv.) and A-FMPC (prop.) for parameter tuning by E-FRIT with two different target trajectories (sinusoidal or square signal) under Case 2 (initial PID gain $\theta=[0.1,\ 0.1,\ 0.001]^T$) in all interval}
  \label{fig:boxplot_all2}
\end{figure}

Finally, the results of a total of 36-case-experiment with design parameter $\lambda$ (nine cases), target trajectory (sinusoidal, square signals) for the pre-experiment, initial PID gains for the pre-experiment (Case 1 and Case 2), and control method (FMPC, A-FMPC) were performed five times each and summarized in box plots in {\bf Figs. \ref{fig:boxplot_ste}} to {\bf \ref{fig:boxplot_all2}}.
{\bf Figures \ref{fig:boxplot_ste}} and {\bf \ref{fig:boxplot_ste2}} show the box plots of the comparison control performance for Case 1 and 2 in the steady-state response ($55$ -$65$ s) using mean absolute error (MAE).
Note that this interval was adopted for the evaluation because the strong asymmetric hysteresis can be observed for the muscle and has the largest oscillatory responses in the control performance.
The cases without box plots show that the system becomes unstable.
From these figures, for Case 1 and values of $\lambda$ greater than 10, it can be observed that MAE are small and achieve good control performance independent of the value of the target trajectory for the prior experiment.
When $\lambda$ is smaller than 5, FMPC is degraded in control performance independent of the value of the target trajectory for the prior experiment.
In particular, for $\lambda=1$ and 2.5, using sin signal for prior experiments, the control performance was significantly degraded.
This is because the time constant of the PL model is optimized to be small due to the small $\lambda$, which generates PID gains to generate excessive inputs.
On the other hand, in Case 2 in {\bf Fig. \ref{fig:boxplot_ste2}}, the control performance of the FMPC is much worse than in Case 1, even for larger values of $\lambda$.
This implies that the optimization of E-FRIT strongly depends on the initial PID gain of the prior experiment, as discussed previously.
In particular, for using sin signal for pre-experiment, the control performance degrades for almost all values of $\lambda$.
Hence, the control performance with oscillatory response cannot be eliminated even if $\lambda$ is increased.
On the other hand, these figures suggest that A-FMPC can achieve high-precision control performance independent of any value of $\lambda$, the target trajectory for the prior experiment, and the initial PID gain for the prior experiment.
In particular, the proposed method is very effective because it allows the designer to safely tune the design parameters since there are no cases of control failure due to instability.

\begin{figure}[t]
\centering
\includegraphics[width=8.5cm]{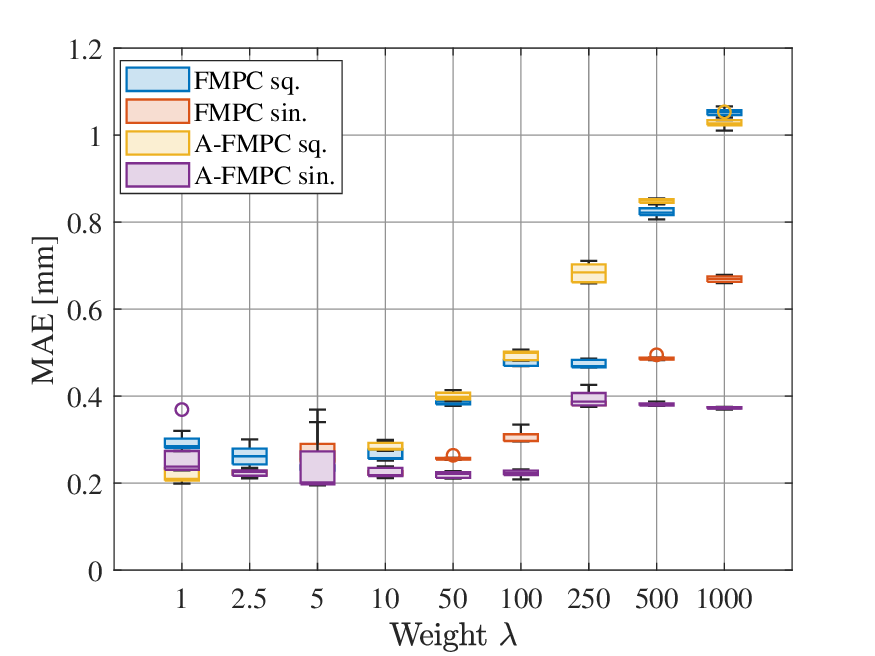}
  \caption{Comparison of box plots for changing the design parameter $\lambda$ between FMPC (conv.) and A-FMPC (prop.) for parameter tuning by E-FRIT with two different target trajectories (sinusoidal or square signal) under Case 1 (initial PID gain $\theta=[0.1,\ 0.1,\ 0.01]^T$) in all interval}
  \label{fig:boxplot_all}
\end{figure}

Moreover, {\bf Figs. \ref{fig:boxplot_all}} and {\bf \ref{fig:boxplot_all2}} show the box plots for Case 1 and 2 in the all interval.
In Case 1, it can be seen that the overall control performance becomes worse as $\lambda$ increases.
This is because the larger the value of $\lambda$, the larger the time constant of the PL model becomes, resulting in a larger deviation between the target trajectory and the reference model response.
A similar tendency is observed in Case 2, but depending on the value of the initial PID gain for the preliminary experiment, the obtained optimal time constant is smaller than in Case 1, and thus has a smaller influence on the control performance.
On the other hand, a too small value of $\lambda$ may destabilize the control system in FMPC.
Therefore, to achieve high control performance for FMPC, it is necessary to select an appropriate $\lambda$.
However, designing $\lambda$ is not easy because it depends on the experimental conditions of the prior experiment and the design of MPC's weights.
On the other hand, the proposed method can achieve sufficiently accurate control as long as the value of $\lambda$ is kept small to prevent deviation between the target trajectory and the reference model response.
Therefore, it is a very practical control method because it can be designed without explicitly a mathematical model and can greatly reduce the trial-and-error process of design.

\newpage

\newpage
\begin{table*}[t]
	\begin{center}
		\caption{Optimized PID gains and time constants by E-FRIT using a staircase square signal for Case 1}
		\label{table:case1}
		\begin{tabular}{cclc}\hline
\rule[0mm]{0mm}{4mm}Case & Design parameter $\lambda$ & Optimal PID gains $\theta^{*}$ & Time constant $T_c$ [s]\rule[0mm]{0mm}{4mm}\\\hline\hline
\rule[0mm]{0mm}{4mm}\multirow{9}{*}{Case 1} & 1.0 & $\theta^{*}=[1.9066,\ 3.4692,\ 9.2513\times 10^{-3}]^T$ & 0.04\\
& 2.5 & $\theta^{*}=[1.4493,\ 2.6192,\ 8.4703\times 10^{-3}]^T$ & 0.05\\
& 5.0 & $\theta^{*}=[1.1763,\ 2.1115,\ 8.321\times 10^{-3}]^T$ & 0.06\\
& 10 & $\theta^{*}=[0.95221,\ 1.6969,\ 8.3838\times 10^{-3}]^T$ & 0.08\\
& 50 & $\theta^{*}=[0.57517,\ 1.0091,\ 8.8442\times 10^{-3}]^T$ & 0.15\\
& 100 & $\theta^{*}=[0.46016,\ 0.80296,\ 9.0583\times 10^{-3}]^T$ & 0.19\\
& 250 & $\theta^{*}=[0.34122,\ 0.59165,\ 9.2993\times 10^{-3}]^T$ & 0.26\\
& 500 & $\theta^{*}=[0.27175,\ 0.46854,\ 9.4413\times 10^{-3}]^T$ & 0.34\\
& 1000 & $\theta^{*}=[0.21663,\ 0.37021,\ 9.5501\times 10^{-3}]^T$ & 0.44\\\hline

\rule[0mm]{0mm}{4mm}\multirow{9}{*}{Case 2}& 1.0 & $\theta^{*}=[2.3612 5.7678 3.6947e\times 10^{-9}]^T$ & 0.02\\
& 2.5 & $\theta^{*}=[1.8034,\ 4.3414,\ 1.9937\times 10^{-9}]^T$ & 0.03\\
& 5.0 & $\theta^{*}=[1.4813,\ 3.5135,\ 1.287\times 10^{-9}]^T$ & 0.03\\
& 10 & $\theta^{*}=[1.22,\ 2.8451,\ 8.4543\times 10^{-8}]^T$ & 0.04\\
& 50 & $\theta^{*}=[0.77497,\ 1.7321,\ 3.9498\times 10^{-10}]^T$ & 0.07\\
& 100 & $\theta^{*}=[0.63441,\ 1.3923,\ 3.3542\times 10^{-10}]^T$ & 0.09\\
& 250 & $\theta^{*}=[0.48427,\ 1.038,\ 4.8044\times 10^{-10}]^T$ & 0.13\\
& 500 & $\theta^{*}=[0.39294,\ 0.82802,\ 7.0176\times 10^{-5}]^T$ & 0.16\\
& 1000 & $\theta^{*}=[0.3176,\ 0.6583,\ 2.415\times 10^{-4}]^T$ & 0.21\\\hline
		\end{tabular}
	\end{center}
\end{table*}

\begin{table*}[t]
	\begin{center}
		\caption{Optimized PID gains and time constants by E-FRIT using a sin signal}
		\label{table:case2}
		\begin{tabular}{cclc}\hline
\rule[0mm]{0mm}{4mm}Case & Design parameter $\lambda$ & Optimal PID gains $\theta^{*}$ & Time constant $T_c$ [s]\rule[0mm]{0mm}{4mm}\\\hline\hline
\rule[0mm]{0mm}{4mm}\multirow{9}{*}{Case 1} & 1.0 & $\theta^{*}=[3.653,\ 3.2734,\ 1.3559\times 10^{-8}]^T$ & 0.02\\
& 2.5 & $\theta^{*}=[2.7955,\ 2.5109,\ 1.1656\times 10^{-3}]^T$ & 0.03\\
& 5.0 & $\theta^{*}=[2.2589,\ 2.0382,\ 3.266\times 10^{-3}]^T$ & 0.03\\
& 10 & $\theta^{*}=[1.8123,\ 1.6455,\ 5.2245\times 10^{-3}]^T$ & 0.04\\
& 50 & $\theta^{*}=[1.0553,\ 0.98224,\ 8.9207\times 10^{-3}]^T$ & 0.07\\
& 100 & $\theta^{*}=[0.82319,\ 0.77986,\ 1.0175\times 10^{-2}]^T$ & 0.10\\
& 250 & $\theta^{*}=[0.58025,\ 0.56908,\ 1.1625\times 10^{-2}]^T$ & 0.14\\
& 500 & $\theta^{*}=[0.43485,\ 0.44366,\ 1.2641\times 10^{-2}]^T$ & 0.18\\
& 1000 & $\theta^{*}=[0.31528,\ 0.34098,\ 1.3673\times 10^{-2}]^T$ & 0.26\\\hline

\rule[0mm]{0mm}{4mm}\multirow{9}{*}{Case 2}& 1.0 & $\theta^{*}=[4.9185,\ 5.1084,\ 8.422\times 10^{-10}]^T$ & 0.01\\
& 2.5 & $\theta^{*}=[4.7233,\ 5.0485,\ 3.4331\times 10^{-10}]^T$ & 0.01\\
& 5.0 & $\theta^{*}=[4.135,\ 4.4636,\ 1.0685\times 10^{-10}]^T$ & 0.01\\
& 10 & $\theta^{*}=[3.5284,\ 3.8318,\ 1.3803\times 10^{-10}]^T$ & 0.01\\
& 50 & $\theta^{*}=[2.4323,\ 2.7184,\ 4.7936\times 10^{-11}]^T$ & 0.02\\
& 100 & $\theta^{*}=[2.0894,\ 2.3977,\ 2.954\times 10^{-11}]^T$ & 0.02\\
& 250 & $\theta^{*}=[1.7336,\ 2.1169,\ 1.505\times 10^{-11}]^T$ & 0.03\\
& 500 & $\theta^{*}=[1.5034,\ 1.9773,\ 8.997\times 10^{-12}]^T$ & 0.03\\
& 1000 & $\theta^{*}=[1.2785,\ 1.8467,\ 1.1113\times 10^{-10}]^T$ & 0.03\\\hline
		\end{tabular}
	\end{center}
\end{table*}

\begin{landscape}
\begin{table}[t]
	\begin{center}
		\caption{Summary of the experimental conditions}
		\label{table:summary}
		\begin{tabular}{cccccclc}\hline
\rule[0mm]{0mm}{4mm}Fig. number & Method & Case & Parameter tuning with & Control for & Design parameter $\lambda$ & Optimized initial PID gain $\theta^{*}$ & Time constant $T_c$ [s]\rule[0mm]{0mm}{4mm}\\\hline\hline

\rule[0mm]{0mm}{4mm}\multirow{4}{*}{\bf Figs. \ref{fig:y_A-FRIT}, \ref{fig:u_A-FRIT}, \ref{fig:u_A-FRIT2}} & A-FRIT & \multirow{4}{*}{Case 1} & \multirow{4}{*}{sq.} & \multirow{4}{*}{sq.} & \multirow{2}{*}{$1$} & \multirow{2}{*}{$\theta^{*}=[1.9066,\ 3.4692, 9.2513\times 10^{-3}]^T$} & \multirow{2}{*}{$0.04$}\\
& A-FMPC &  &  &  &  & & \\
& A-FRIT &  &  &  & \multirow{2}{*}{$100$} & \multirow{2}{*}{$\theta^{*}=[0.46016,\ 0.80296, 9.0583\times 10^{-3}]^T$} & \multirow{2}{*}{$0.19$}\\
& A-FMPC &  &  &  &  &  & \\\hline

\rule[0mm]{0mm}{4mm}\multirow{4}{*}{\bf Figs. \ref{fig:y_FMPC}, \ref{fig:me_FMPC}} & FMPC & \multirow{4}{*}{Case 1} & \multirow{2}{*}{sq.} & \multirow{4}{*}{sq.} & \multirow{4}{*}{$100$} & \multirow{2}{*}{$\theta^{*}=[0.46016,\ 0.80296, 9.0583\times 10^{-3}]^T$} & \multirow{2}{*}{$0.19$}\\
& A-FMPC &  &  &  &  & & \\
& FMPC &  & \multirow{2}{*}{sin.} &  &  & \multirow{2}{*}{$\theta^{*}=[0.82319\ ,0.77986,\ 1.0175\times 10^{-2}]^T$} & \multirow{2}{*}{$0.10$}\\
& A-FMPC &  &  &  &  &  & \\\hline

\rule[0mm]{0mm}{4mm}\multirow{4}{*}{\bf Figs. \ref{fig:y_FMPC2}, \ref{fig:me_FMPC2}} & FMPC & \multirow{4}{*}{Case 2} & \multirow{2}{*}{sq.} & \multirow{4}{*}{sq.} & \multirow{4}{*}{$100$} & \multirow{2}{*}{$\theta^{*}=[0.63441,\ 1.3923, 3.3542\times 10^{-10}]^T$} & \multirow{2}{*}{$0.09$}\\
& A-FMPC &  &  &  &  & & \\
& FMPC &  & \multirow{2}{*}{sin.} &  &  & \multirow{2}{*}{$\theta^{*}=[2.0894\ ,2.3977,\ 2.954\times 10^{-11}]^T$} & \multirow{2}{*}{$0.02$}\\
& A-FMPC &  &  &  &  &  & \\\hline
		\end{tabular}
	\end{center}
\end{table}
\end{landscape}

\clearpage

\section{Conclusion}
A novel data-driven adaptive model matching-based model predictive control for water-hydraulic artificial muscles is proposed.
The proposed method achieved an improvement of the each disadvantage of the conventional method, namely, the dependence of FMPC on prior experimental data and the degradation of transient performance due to excessive overshooting of A-FRIT with directional forgetting.
In this paper, it is shown that the conventional method achieves the same control performance as the proposed method, but the control performance is significantly degraded under some conditions.
For a more detailed evaluation, experimental conditions covering as many as 36 cases that the proposed method achieves high robustness concerning the values of the target trajectory and the initial PID gain for the pre-experimental data, and the design parameter $\lambda$ of the E-FRIT.
In addition, the results of the analyses provide guidelines for the design of the A-FMPC.
Furthermore, these results cloud be achieved without the explicit use of mathematical models, and the water-hydraulic artificial muscles have a very low environmental impact because they do not require a special power source and are completely oil-free, so they are expected to be applied in various fields.

\acknowledgements
This work was supported by JSPS Grant-in-Aid for JSPS Fellows Grant Number JP23KJ1923.

\end{document}